\documentclass[10pt, conference, letterpaper]{IEEEtran}
\usepackage{booktabs} 
\usepackage{enumitem} 
\usepackage{cleveref}
\usepackage[english]{babel}
\usepackage{import}
\usepackage{subfiles}
\usepackage{array}
\usepackage{makecell}
\usepackage{booktabs}
\usepackage{multirow}
\usepackage{hhline}
\usepackage{makecell}
\usepackage[font=footnotesize]{subcaption}
\usepackage[font=footnotesize]{caption}
\usepackage{csquotes}

\usepackage{cite}

\ifCLASSINFOpdf
  \usepackage[pdftex]{graphicx}
  \graphicspath{{figures/}{../figures/}}
  \DeclareGraphicsExtensions{.pdf,.jpeg,.png}
\else
  \usepackage[dvips]{graphicx}
  \graphicspath{{figures/}}
  \DeclareGraphicsExtensions{.eps}
\fi
\usepackage{amsmath}
\usepackage{array}
\usepackage{url}

\begin{document}
%
\title{EVSO: Environment-aware Video Streaming Optimization of Power Consumption}

\author{\IEEEauthorblockN{Kyoungjun Park and Myungchul Kim}
\IEEEauthorblockA{School of Computing\\
Korea Advanced Institute of Science and Technology\\
Email: \{kyoungjun, mck\}@kaist.ac.kr}}


%


\maketitle

\begin{abstract}
Streaming services gradually support high-quality videos for the better user experience. However, streaming high-quality video on mobile devices consumes a considerable amount of energy. This paper presents the design and prototype of EVSO, which achieves power saving by applying adaptive frame rates to parts of videos with a little degradation of the user experience. EVSO utilizes a novel perceptual similarity measurement method based on human visual perception specialized for a video encoder. We also extend the media presentation description, in which the video content is selected based only on the network bandwidth, to allow for additional consideration of the user's battery status. EVSO's streaming server preprocesses the video into several processed videos according to the similarity intensity of each part of the video and then provides the client with the processed video suitable for the network bandwidth and the battery status of the client's mobile device. The EVSO system was implemented on the commonly used H.264/AVC encoder. We conduct various experiments and a user study with nine videos. Our experimental results show that EVSO effectively reduces the energy consumption when mobile devices uses streaming services by 22\% on average and up to 27\% while maintaining the quality of the user experience.
\end{abstract}


%
\IEEEpeerreviewmaketitle

\section{Introduction}\label{sec:introduction}


Video streaming on mobile devices such as smartphones and tablets has seen unprecedented growth in recent years. According to \cite{CiscoWhitePaper}, mobile video traffic accounted for more than 60\% of the total mobile traffic in 2017, and it is expected to grow by 54\% annually, reaching 78\% of all traffic by 2021. Currently, many video streaming services and smartphones support high frame rates and high resolutions for better user experiences \cite{YouTubeRecommend, SamsungSmartphones}. In addition, both Virtual Reality (VR) and Augmented Reality (AR) that require high-quality video are considered to be among the next big applications in mobile technology. 
However, high-quality videos require a significant proportion of device resources, mainly display-related components, resulting in much higher power consumption \cite{lim2016adaptive, RAVEN}.

Many mobile device manufacturers have put in much effort to increase the battery lifetime of mobile devices. According to Samsung, the battery capacity of their smartphones has experienced a Compound Annual Growth Rate (CAGR) of 11.17\% from 2010 to present indicating that the battery capacity has steadily improved \cite{SamsungSmartphones}. However, since the amount of mobile video traffic will increase by 54\% yearly, the current rate of improvement in the battery capacity will not meet the power consumption requirement of video streaming in the near future.

There have been several efforts to reduce the power consumption of mobile games, which typically have high frame rates and resolutions causing the rapid battery drain. One approach called Game Tuner \cite{GameTuner} allows users to configure parameters related to frame rates or resolutions. However, the disadvantages of this approach are that the user experience is seriously degraded due to the static configuration and that the user intervention is required. To solve these problems, there have been several approaches that attempt to dynamically scale frame rates based on the frame contents \cite{RAVEN, kim2016content}. This is achieved by measuring the structural similarity between frames and then dropping redundant or very similar frames. However, these approaches only drop frames on the client side; hence, the number of frames transmitted to the client is not affected. Therefore, the energy consumption required for wireless transmission on the client side remains the same. In addition, the network bandwidth on the client side is not utilized efficiently because unnecessary frames are transmitted.


Based on the aforementioned discussion, we propose Environment-aware Video Streaming Optimization (EVSO), which effectively applies adaptive frame rates for videos without requiring any user effort or incurring considerable computational overhead. 
The environment represents the status of the user's mobile device and the characteristics of the video that the streaming server delivers to the user.
The basic idea behind EVSO is that not all parts of the video require high frame rates. In other words, adaptive frame rates can be applied to parts of the video according to the degree of motion change between frames. For example, in a video where a golfer prepares for a swing, a lower frame rate can be applied due to the low variation in the frames \cite{ExperimentVideos}. On the other hand, a higher frame rate can be applied to the high variation in the frames such as a swing in motion or a moving golf ball to avoid a degradation of the user experience.

EVSO utilizes a H.264/AVC encoder \cite{H264Overview}, which is the most widely used video encoder, to be interoperable with existing systems without additional deployment overhead. EVSO includes the \textit{Frame rate Scheduler} (F-Scheduler), which scales the frame rates of specific parts of videos using a perception-aware analysis based on the information generated during the H.264/AVC encoding process. We note that the H.264/AVC encoder calculates the similarity between frames based on \textit{macroblocks} during the video compression (or video coding) process. 
EVSO utilizes these macroblocks to schedule different parts of the video with appropriate frame rates.

The EVSO system was implemented on a video streaming server built with Internet Information Services (IIS) \cite{IIS}. We conducted a broad range of experiments and a user study with nine videos to evaluate the proposed system. Our experimental results show that EVSO can greatly reduce the energy requirement with little impact on the user experience. The system reduced the energy consumption rate by an average of 22\% and up to 27\%. In addition, the user study shows that users cannot clearly distinguish between the original videos and videos processed through EVSO.

The main contributions of this paper are as follows:
\begin{itemize}[noitemsep,topsep=0pt]
  
  \item We propose a new perceptual similarity calculation method that leverages the information generated by the H.264/AVC encoding process. 
  
  \item Based on the similarity calculation method, we present a novel scheduling technique that adaptively adjusts the frame rate according to the degree of motion intensity.
  
  \item We extend the media presentation description (MPD) to take into account not only network conditions but also battery status.
  
  \item We present the design and prototype of the EVSO system to reduce energy consumption when streaming videos.
  
  \item Various experiments and a user study show that the proposed system effectively preserves the user experience and reduces the power consumption when streaming videos.
  
\end{itemize}

The rest of this paper is organized as follows. Section~\ref{sec:motivation} explains the motivation behind EVSO. Section~\ref{sec:design} describes the design of EVSO and its main functions. Section~\ref{sec:implementation} presents the detailed implementation of EVSO and Section~\ref{sec:evaluation} reports the various experiments and a user study. 
Section~\ref{sec:related} introduces the related work and Section~\ref{sec:conclusion} concludes the study and discusses possible future work.

\section{Motivation and Background}\label{sec:motivation}

This section highlights the necessity of EVSO and explains the information leveraged by EVSO during the H.264/AVC encoding process.

\subsection{Similarity Analysis of Adjacent Video Frames}
\begin{figure}[!t]
    \includegraphics[width=1\columnwidth]{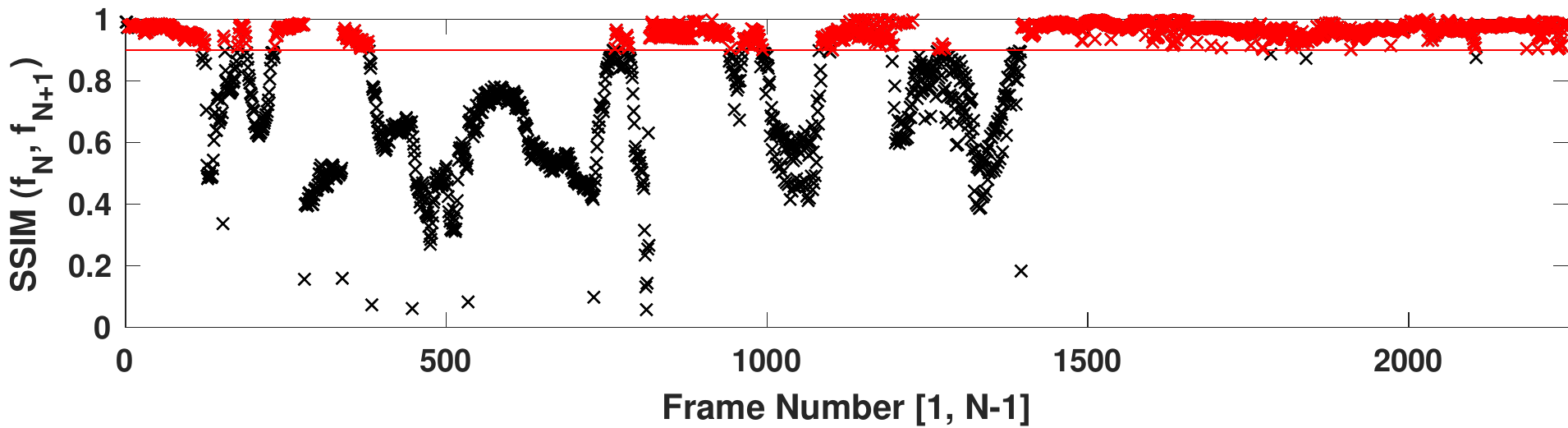}
    \caption{Variation of SSIM between adjacent frames in the baseball video.}
    \label{fig:SSIMBaseball}
    \vspace{-1.0em}
\end{figure}
\begin{figure}[!t]
    \includegraphics[width=1\columnwidth]{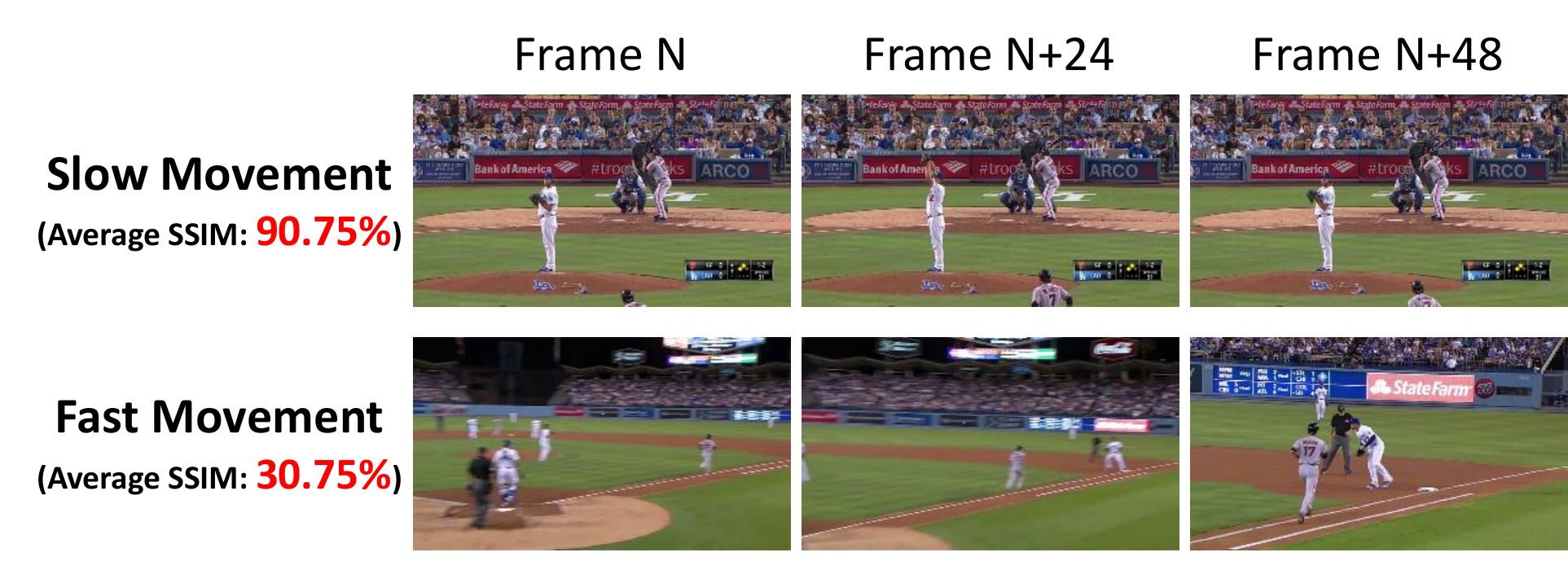}
    \caption{Degree of frame change during fast and slow movements.}
    \label{fig:FrameChange}
    \vspace{-1.0em}
\end{figure}

The similarity between frames can be measured using the Structural SIMilarity (SSIM) index \cite{SSIM}. Figure~\ref{fig:SSIMBaseball} shows SSIM indices between adjacent frames in a baseball video \cite{ExperimentVideos}. The range of the SSIM index is from 0 to 1, and a value of 0.9 or higher is considered to indicate a strong similarity between two frames \cite{Kahawai}. The red shaded mark in Figure~\ref{fig:SSIMBaseball} indicates that the SSIM value between adjacent frames is higher than 0.9, which means that the degree of change is fairly low. For example, Figure~\ref{fig:FrameChange} shows examples of a slow movement where the pitcher prepares to throw the ball and a fast movement where the ball is thrown and the camera quickly follows the ball. The difference between the average values of SSIM for the two movement scenarios is 60\%, which is a large gap. This indicates that the degree of change can vary greatly according to which part of the video is playing, even within in a single video. Therefore, the energy consumption requirement can be reduced by throttling down the frame rate during slow moving parts.

\subsection{Video Upload Process and H.264/AVC Encoder}
When a user uploads a video to a streaming media service such as YouTube, the server \textit{transcodes} the video into a set of optimized video streams. Transcoding is a conversion process in which the encoding settings of the original video are reformatted. There are several reasons why the streaming server must perform transcoding when the video is uploaded. 
The first case is when the streaming server needs to lower the bitrate of the uploaded video, which is called \textit{transrating}. The second case is when the streaming server needs to lower the resolution of the uploaded video, known as \textit{transsizing}. The transrating and transsizing are always performed when the server creates several processed videos to provide DASH functionality \cite{DASH}.

The most commonly used video codec for streaming servers is H.264/AVC \cite{YouTubeRecommend}. Therefore, when transcoding is performed on a streaming server, the H.264/AVC encoder is used to convert the original video into the desired videos. The video encoding of H.264/AVC follows a block-based approach where each coded picture is represented by block-shaped units called macroblocks. A macroblock consists of four 8\(\times\)8 luminance (Y) samples and two 8\(\times\)8 chrominance (Cb and Cr) samples in the YCbCr color space with 4:2:0 chroma subsampling \cite{H264Overview}. 
These Y samples are extracted and utilized in EVSO because the human visual system is sensitive to luminance \cite{Vision}. 

\section{EVSO Design}\label{sec:design}



\begin{figure}[!t]
    \includegraphics[width=1\columnwidth]{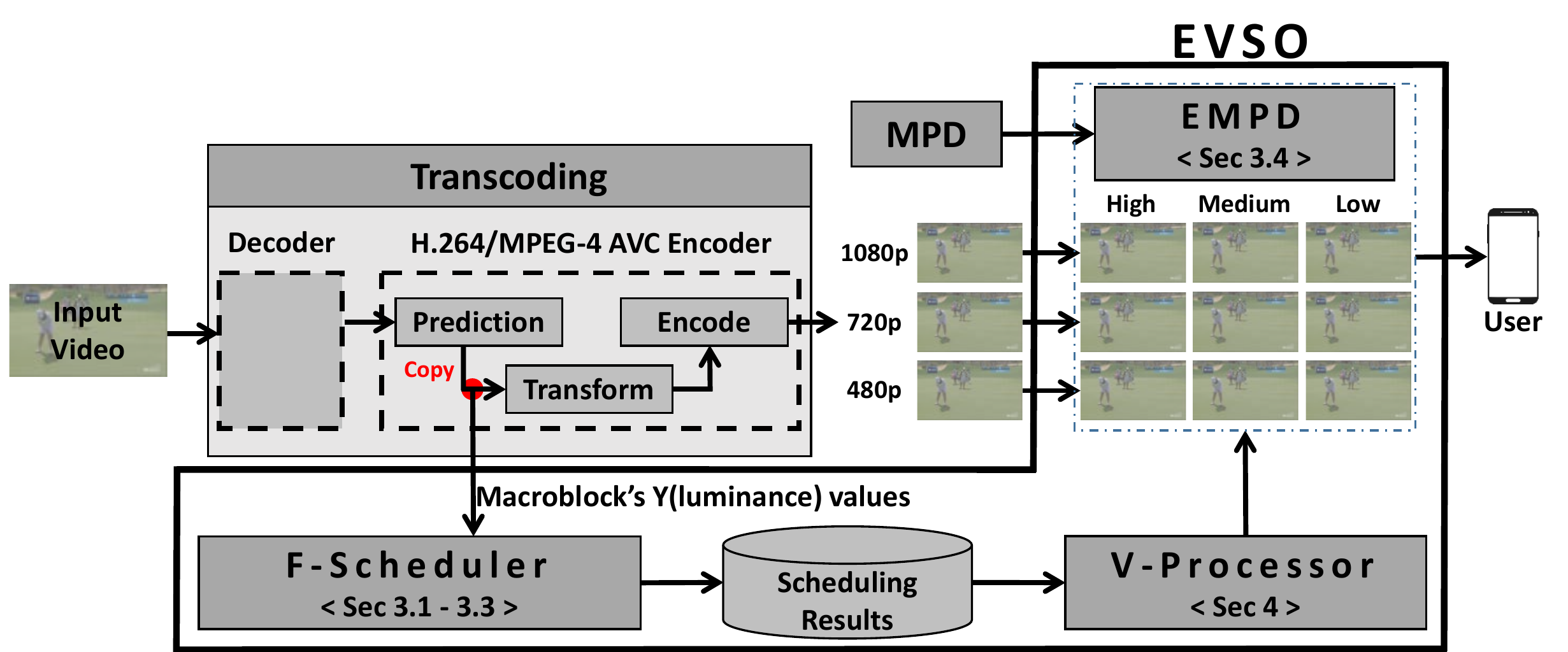}
    \caption{Flow of EVSO when a video is uploaded to a streaming server.}
    \label{fig:EVSOArchitecture}
    \vspace{-1.0em}
\end{figure}

Figure~\ref{fig:EVSOArchitecture} shows the EVSO system, which consists of three major components: \textit{Frame rate Scheduler} (F-Scheduler), \textit{Video Processor} (V-Processor), and \textit{Extended MPD} (EMPD). When a video is uploaded to a streaming server, F-Scheduler calculates the perceptual similarity score between adjacent frames with little computational overhead (see Section~\ref{subsec:calculating}). Then, F-Scheduler determines how to split the video into multiple video chunks with similar motion intensity levels (see Section~\ref{subsec:splitting}) and schedules the appropriate frame rates for each video chunk according to three battery levels: High, Medium, and Low (see Section~\ref{subsec:estimating}). Although our study adopts three battery levels, other settings can also be applied. The scheduling results generated from F-Scheduler specify which frame rate is appropriate for each part of the video. Afterwards, V-Processor processes the original video according to what is specified in the scheduling results to produce videos suitable for the three battery levels. Detailed implementation of V-Processor is described in Section~\ref{sec:implementation}. Finally, EMPD allows users to request appropriate video chunks, taking into consideration not only the network condition but also the battery status (see Section~\ref{subsec:extending}). 

Note that this paper focuses on H.264/AVC as the video codec; however, we emphasize that the proposed EVSO system is not limited solely to H.264/AVC. Since other codecs also perform video compression using the luminance (Y) values of the macroblock in the similar manner as H.264/AVC, EVSO can easily be applied to other codecs. 



\subsection{Calculating the Perceptual Similarity Score}\label{subsec:calculating}
The human vision system is more sensitive to changes in the brightness (Y) than it is to changes in colors (Cb and Cr) \cite{Vision, H264Overview}. Hwang et al. \cite{RAVEN} have shown that the Y-Difference (or Y-Diff) reflects the changes in visual perception fairly well. Moreover, they found Y-Diff to be highly correlated with the commonly used SSIM method through various experiments and a user study. 
However, human vision is sensitive not only to brightness but also to object movement \cite{Sensation}. The Y-Diff value between two images is calculated as the Sum of Absolute Differences (SAD) of the luminance values on a per-frame basis, not on a per-block basis. Therefore, the local information can be ignored because the detailed features of each block are mixed. In other words, Y-Diff effectively reflects the brightness but can be less accurate to the motion perception of objects. Although Y-Diff can partially reflect some perception of object movement \cite{Vision}, but more attention needs to be paid to motion information in order to be more compatible with human perception.

To effectively calculate the perceptual similarity scores by considering both the brightness and object movement in video streaming environments, we define the \textit{Macroblock-Difference} (or M-Diff), \( D_{MB}(f_{a}, f_{b})\), between frames \(f_{a}\) and \(f_{b}\) as follows:

\begin{footnotesize}
\begin{equation}
    \label{D_MB}
    D_{MB}(f_{a}, f_{b}) = \sum_{i=0}^{N-1}\sum_{j=0}^{M-1} D_{Y}(f_{a}(i, j), f_{b}(i, j))
\end{equation}
\begin{equation}
    \label{D_Y}
    \resizebox{0.9\columnwidth}{!}{$
    D_{Y}(f_{a}(i, j), f_{b}(i, j)) =
    \begin{cases}
        1, & SAD_{Y}(f_{a}(i, j), f_{b}(i, j)) > \theta \\
        0, & SAD_{Y}(f_{a}(i, j), f_{b}(i, j)) \leq \theta
    \end{cases}
    $}
\end{equation}
\end{footnotesize}

\noindent where \(f_{k}(i, j)\) is the \(k\)-th frame of \(N \times M\) macroblocks with \(i\) and \(j\) representing the macroblock coordinates; \(SAD_{Y}(f_{a}(i, j), f_{b}(i, j))\) is the SAD value between the luminance value of \(f_{a}(i, j)\) and \(f_{b}(i, j)\); and \(D_{Y}(f_{a}(i, j), f_{b}(i, j))\) is set to one when the SAD value between the luminance value of \(f_{a}(i, j)\) and \(f_{b}(i, j)\) exceeds a certain threshold \(\theta\). The value of \(\theta\) is set to 320 because OpenH264 \cite{OpenH264} determines high motion blocks when the SAD value of the luminance exceeds 320 when detecting a scene change. Finding the optimal value of \(\theta\) is left as future work.

\begin{figure}[!t]
    \includegraphics[width=1\columnwidth]{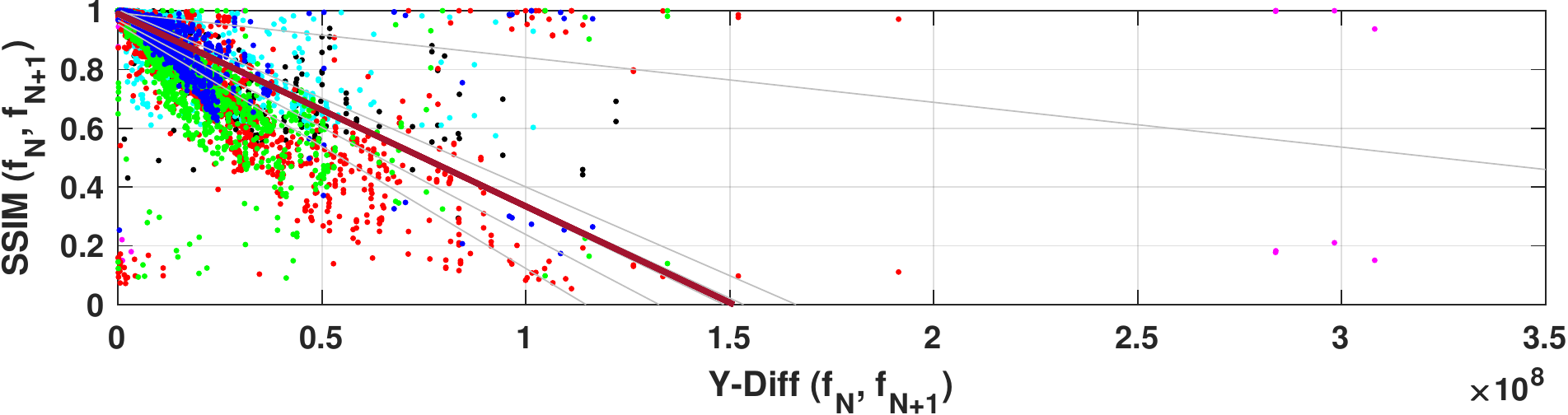}
    \caption{Correlation between Y-Diff and SSIM (Pearson's correlation coefficient: -0.7329).}
    \label{fig:YDiff_corr}
    \includegraphics[width=1\columnwidth]{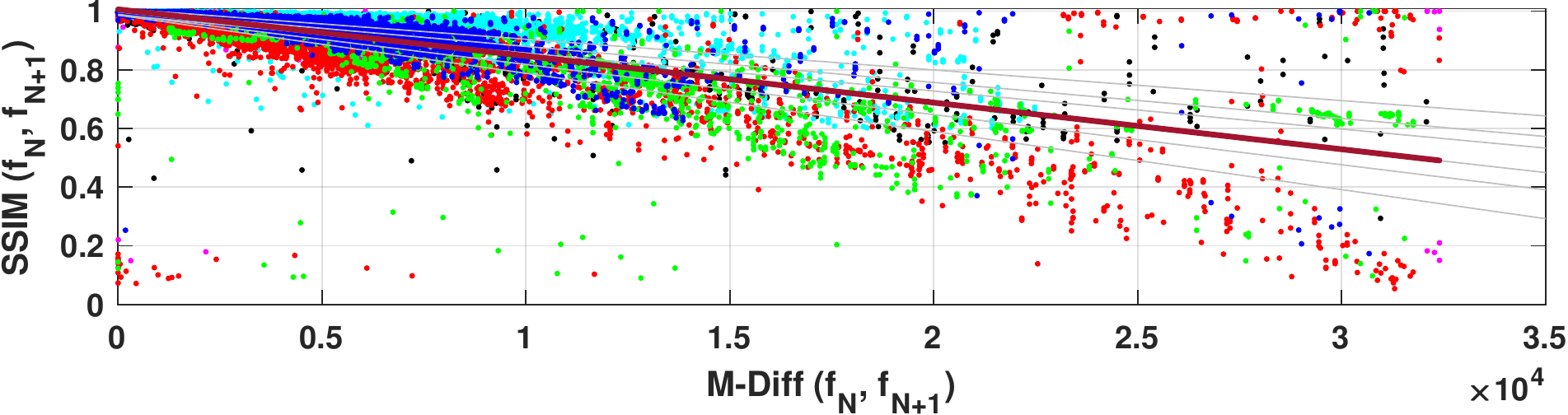}
    \caption{Correlation between M-Diff and SSIM (Pearson's correlation coefficient: -0.7834).}
    \label{fig:MDiff_corr}
    \vspace{-1.0em}
\end{figure}

EVSO used M-Diff as a perceptual similarity measurement method to cover both brightness and object movement. M-Diff uses the luminance values of macroblocks to perform brightness perception and handles object movement through block-based calculation. The block-based approach performs the perceptual similarity calculation based on the macroblock unit rather than on the frame unit. Figure~\ref{fig:YDiff_corr} shows how Y-Diff is similar to SSIM in the analysis on adjacent frames of seven videos \cite{ExperimentVideos}, and Figure~\ref{fig:MDiff_corr} shows how M-Diff is similar to SSIM in the same videos. These figures indicate that M-Diff is highly correlated with SSIM, which is approximately 5\% higher than Y-Diff. In other words, M-Diff has an accuracy comparable to or better than Y-Diff in perceptual similarity calculation.

\noindent \textbf{How about using SSIM as a method for the perceptual similarity calculation:}
More than 72 hours of new video content is uploaded to YouTube every minute, and the volume of uploaded videos continues to grow \cite{YouTubeUploadAmount}. For such a large number of videos, performing SSIM calculations for consecutive frames is exorbitant in terms of computational time and power. On a desktop PC equipped with a 3.5 GHz processor and 12 GB of memory, it took about 401 milliseconds to calculate the perceptual similarity between two frames with 1920\(\times\)1080 resolution using SSIM. If SSIM is used to process a video with 60 frames per second for a duration of 30 minutes, the time overhead is approximately 12 hours.
On the other hand, there is little or no overhead incurred in the M-Diff calculation because it uses the luminance differences between macroblocks generated by the H.264/AVC encoder when a video is uploaded.


\subsection{Splitting Video into Multiple Video Chunks}\label{subsec:splitting}

In order to apply multiple frame rates to a single video, how to properly split the video into multiple chunks needs to be considered. If a video is split too finely, an appropriate frame rate can be set to suit the characteristics of each video chunk, but it takes a very long time to process the video. In contrast, if a video is split too coarsely, it is difficult to calculate the appropriate frame rate because the characteristic of each video chunk becomes ambiguous, which can adversely impact the user experience.

\noindent \textbf{Estimating the split threshold:}
A simple approach to splitting the video is to use the M-Diff value in the perceptual similarity score as a separation criterion. For example, if more than 80\% of the corresponding macroblocks in two frames exceed the threshold \(\theta\), then there is a large enough difference between the two frames and thus they can be separated. 
Another approach is to use the local statistical properties of the frame sequence to define a dynamic threshold model. These local properties can include mean and standard deviation to determine the degree of change in the frame sequence. However, the above two approaches are used mainly to detect scene changes \cite{OpenH264}. On the other hand, our approach is to design separation criteria that allow each video chunk to have a similar degree of variability.

We newly define the \textit{Estimated Split Threshold} (EST) to separate the video into multiple chunks with similar variability levels, as follows:

\begin{footnotesize}
\begin{equation}
\resizebox{0.9\columnwidth}{!}{$
    EST(f_{n}) =
    \begin{cases}
        1, & \text{if } (\sigma_{n} > \alpha\ \text{or}\ D_{MB}(f_{n-1}, f_{n}) > \beta)\ \text{and}\ T > \gamma \\
        0, & otherwise
    \end{cases}
$}
\end{equation}
\begin{equation} \label{eq:sigma_n}
    \sigma_{n} = \sqrt{\frac{1}{K-1} \sum_{i=n-K+1}^{n-1} (D_{MB}(f_{i}, f_{i+1})-m_{n})^2}
\end{equation}
\begin{equation}
    m_{n} = \frac{1}{K} \sum_{i=n-K+1}^{n-1} D_{MB}(f_{i}, f_{i+1})
\end{equation}
\end{footnotesize}

\noindent where \(m_{n}\) is the average M-Diff value of the previous \(K\) number of frames; \(\sigma_{n}\) is the standard deviation with the window size \(K\) representing the degree of M-Diff scattering of the previous \(K\) number of frames; and \(EST(f_{n})\) is a threshold that determines whether to split. When the value of \(EST(f_{n})\) is one, EVSO decides to split at \(f_{n}\). In \(EST(f_{n})\), \(D_{MB}(f_{n-1}, f_{n})\) is used to detect clear scene changes. Finally, \(T\) is the number of frames in the current video chunk and \(\gamma\) is the frame rate of the current video. The values of \(\sigma_{n}\) and \(D_{MB}(f_{n-1}, f_{n})\) tend to be large around the high-motion part of the video, which can cause the video to split too finely. Therefore, the inequality \(N > \gamma\) representing the minimum condition is applied so that each video chunk is at least one second long.

\noindent \textbf{Determining the \(\alpha\), \(\beta\), and \(K\) factors:}
\begin{figure}[!t]
\centering
    \begin{subfigure}[b]{0.45\columnwidth}
        \includegraphics[width=\columnwidth]{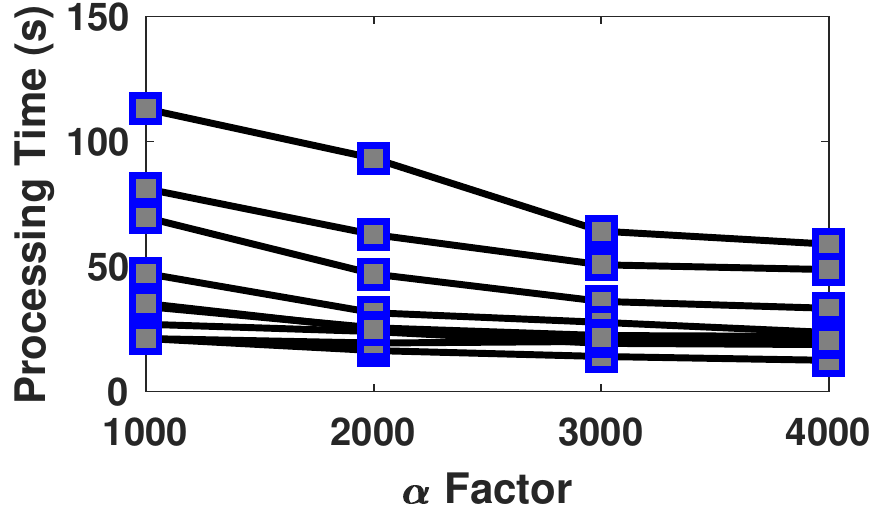}
        \caption{Constant factor \(\alpha\)}
        \label{fig:a_processing_time}
    \end{subfigure}
    \begin{subfigure}[b]{0.45\columnwidth}
        \includegraphics[width=\columnwidth]{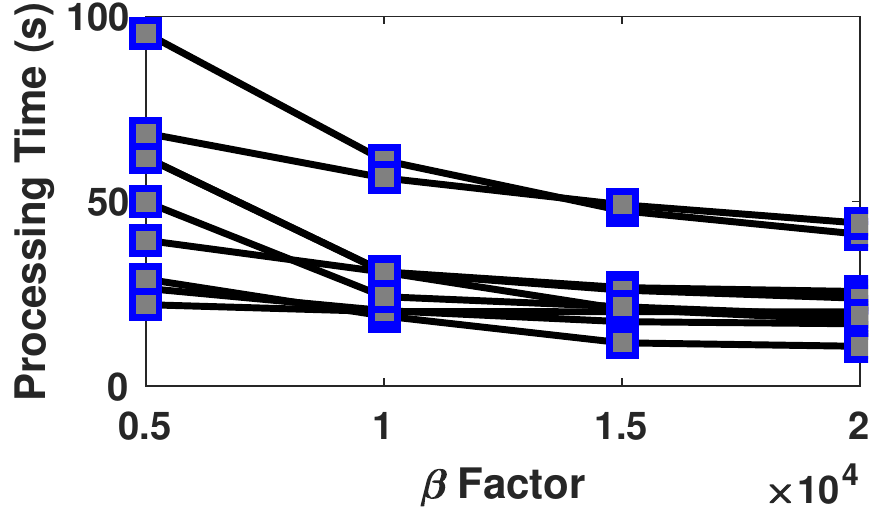}
        \caption{Constant factor \(\beta\)}
        \label{fig:b_processing_time}
    \end{subfigure}
    \caption{Variation of the processing time according to the constant factors.}
    \label{fig:processing_time}
    \vspace{-1.0em}
\end{figure}
The smaller the value of the constant factors \(\alpha\) and \(\beta\), the less frames are assigned to the video chunk, so each video chunk can have a more similar level of variability. In the extreme, setting \(\alpha\) and \(\beta\) to 0 can be effective for the user experience; however, if \(\alpha\) and \(\beta\) are too low, the number of separated video chunks and the processing time of V-Processor increase exponentially. That is, \(\alpha\) and \(\beta\) should be appropriately selected in consideration of the tradeoff between the computational complexity and the accuracy of the frame rate estimation process. 

Figure~\ref{fig:processing_time}(a) shows the processing time when \(\alpha\) is set to 1000, 2000, 3000, and 4000 with nine experimental videos on a server equipped with 2.20 GHz\(\times\)40 processors and 135 GB of memory. As \(\alpha\) increases, the total number of video chunks to be separated decreases reducing the overall processing time. Our findings indicate that in most videos, the processing time decreases drastically when \(\alpha\) is set to 3000, after which it remains nearly the same. Likewise, Figure~\ref{fig:processing_time}(b) shows that the processing time decreases as \(\beta\) increases. Moreover, in most of the videos, the processing time decreases sharply until \(\beta\) reaches 15000 and then remains almost the same. Consequently, \(\alpha\) and \(\beta\) are set to 3000 and 15000, respectively, to allow efficient processing.

If the window size \(K\) is set too high, rapid motion changes cannot be detected as separation criteria. \(K\) values of 1, 5, 10, and 30 were tested to determine the appropriate value for detecting motion changes in our experimental environment. Based on numerous experiments, \(K\) is set to 10. The optimal value of the window size \(K\) depends on the characteristic of the video being processed.

\subsection{Estimating Frame Rates for Video Chunks}\label{subsec:estimating}
After splitting the video into multiple video chunks based on the discussion presented in Section~\ref{subsec:splitting}, an appropriate frame rate needs to be determined for each video chunk. Setting the proper frame rate to match the variation of each video chunk is a crucial step in terms of the user experience.

\begin{figure*}[!t]
    \includegraphics[width=0.95\textwidth]{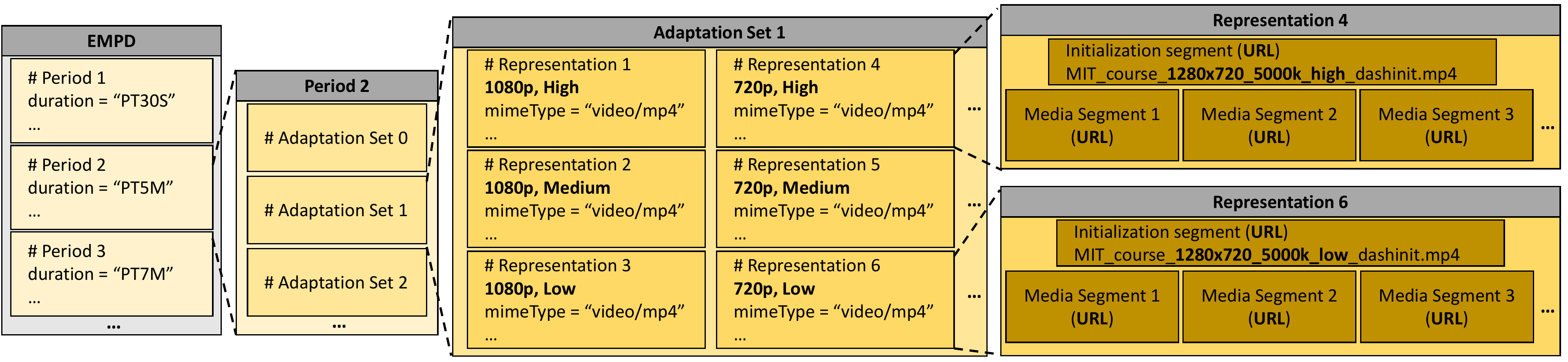}
    \caption{The overall hierarchical structure of EMPD.}
    \label{fig:EMPD}
    \vspace{-1.0em}
\end{figure*}
\noindent \textbf{Estimating proper frame rates for video chunks:} Based on the analysis of the relationship between the M-Diff values and frame rates, we define the \textit{Estimated Proper Frame rate} (EPF) as follows:

\begin{footnotesize}
\begin{equation} \label{eq:EPF}
    EPF(f_{n}, f_{n+1}) =
    \begin{cases}
        s_{1} \times \gamma & \quad \text{if } D_{MB}(f_n, f_{n+1}) < \tau_{1} \\
        s_{2} \times \gamma & \quad \text{if } \tau_{1} \leq D_{MB}(f_n, f_{n+1}) < \tau_{2} \\
        s_{3} \times \gamma & \quad \text{if } \tau_{2} \leq D_{MB}(f_n, f_{n+1}) < \tau_{3} \\
        s_{4} \times \gamma & \quad \text{if } \tau_{3} \leq D_{MB}(f_n, f_{n+1}) < \tau_{4} \\
        s_{5} \times \gamma & \quad \text{if } \tau_{4} \leq D_{MB}(f_n, f_{n+1})
    \end{cases}
\end{equation}
\end{footnotesize}

\noindent where \(s_{1}\), \(s_{2}\), \(s_{3}\), \(s_{4}\), and \(s_{5}\) are the scaling factors of the frame rate \(\gamma\) that match the degree of change between \(f_{n}\) and \(f_{n+1}\); and \(\tau_{1}\), \(\tau_{2}\), \(\tau_{3}\), and \(\tau_{4}\) are the scaling thresholds that distinguish M-Diff values based on similar variations. \(EPF(f_{n}, f_{n+1})\) is utilized to calculate the appropriate frame rate for adjacent frames, and then we define the \textit{Estimated Video chunk Frame rate} (EVF) to obtain the appropriate frame rate for each video chunk as follows:

\begin{footnotesize}
\begin{equation} \label{eq:EVF}
    EVF(S_{k}) = \sum_{i=l}^{m-1} \frac{EPF(f_{i}, f_{i+1})}{m-l} + \delta \times \sigma_{S_{k}}
\end{equation}
\end{footnotesize}

\noindent where \(S_{k}\) represents the \(k\)-th video chunk, i.e., \((f_{l}, f_{l+1}, f_{l+2}, \ldots, f_{m})\), which is needed to estimate the appropriate frame rate;  
and \(\sigma_{S_{k}}\) is the standard deviation of adjacent frames in \(S_{k}\) based on Equation~(\ref{eq:sigma_n}), which is used to determine whether the degree of M-Diff variation in the video chunk is constant or anomalous. If \(\sigma_{S_{k}}\) is high, it means that fast and slow moving parts coexist in a single video chunk. In this case, the overall frame rate can be adjusted to take into account the highly variable parts for a better user experience. Our experiments were performed with \(\delta\) equal to 0.0001, but it can be tailored to the specific environment in which EVSO is deployed.

\noindent \textbf{Determining \(\tau\) and \(s\) factors:}
To determine the \(\tau\) factors, we analyzed the relationship between M-Diff and the commonly used SSIM using a linear regression based on Figure~\ref{fig:MDiff_corr}. The resulting equation is as follows:

\begin{footnotesize}
\begin{equation} \label{eq:regression}
    SSIM(f_{n}, f_{n+1}) = 1.0063 - 1.5903 \times 10^{-5} \times D_{MB}(f_{n}, f_{n+1}).
\end{equation}
\end{footnotesize}

The statistical values of \(R^{2}\) (R-squared) and Pearson Correlation Coefficient (PCC) were calculated to measure the goodness of fit. The \(R^{2}\) and PCC values of this regression model are 0.613 and -0.7834, respectively, indicating that the correlation between SSIM and M-Diff is sufficiently high.

If the SSIM index between two frames is above 0.9, the peak signal-to-noise ratio is above 50 dB and the mean opinion score, which represents the quality of experience on a scale of 1 to 5, corresponds to 4 (Good) or 5 (Excellent, Identical) \cite{Kahawai}. Based on this, we subdivided the SSIM index into 0.99, 0.98, 0.95, and 0.91. According to the regression model in Equation~(\ref{eq:regression}), the M-Diff values corresponding to these SSIM indices are 500, 1500, 3000, and 6000, and these values represent \(\tau_{1}\), \(\tau_{2}\), \(\tau_{3}\), and \(\tau_{4}\) in Equation~(\ref{eq:EPF}), respectively.

The \(s\) factors are set differently depending on the battery level. Detailed settings of the \(s\) factors are described in Section~\ref{sec:evaluation}.

\subsection{Extending Media Presentation Description (MPD) for User Interaction}\label{subsec:extending}
Dynamic Adaptive Streaming over HTTP (DASH) \cite{DASH}, also known as MPEG-DASH, is an adaptive bitrate streaming technology that allows a multimedia file to be partitioned into several segments and transmitted to the client over HTTP. In other words, when a video is uploaded, the streaming server must generate several processed videos by adjusting the bitrate and resolution to provide DASH functionality. 

Media Presentation Description (MPD) refers to a manifest 
that provides segment information such as resolutions, bitrates, and URLs of the video contents. 
Therefore, the DASH client chooses the segment with the highest possible resolution and bitrate that can be supported by the network bandwidth and then fetches the corresponding segment through MPD. EVSO extends MPD to allow the client to select segments considering the battery status as well as the network bandwidth.

Figure~\ref{fig:EMPD} shows the hierarchical structure of \textit{Extended MPD} (EMPD), which consists of one or more \textit{periods} that describe the content duration. Multiple periods can be used when it is necessary to classify multiple videos into chapter-by-chapter or separate advertisements and contents. Each period consists of one or more \textit{adaptation sets}, which include media streams. The period typically consists of separate adaptation sets of audio and video for efficient bandwidth management. The adaptation set of the video consists of several \textit{representations} that contain specific information about alternative contents, such as the resolution and MIME type (or content type). EMPD extended this adaptation set to include an \textit{EVSOLevel} attribute, allowing the client to select the representation that is appropriate for their current battery situation. Finally, the representation provides the URL so that the client can fetch the media segments and play them back.

\section{Implementation}\label{sec:implementation}

In order to study the effectiveness of EVSO, we implemented a video streaming server with DASH functionality using IIS \cite{IIS}. The video streaming server sends EMPD to the client to allow the users to select the appropriate video. We also implemented a video streaming client by modifying ExoPlayer \cite{GithubExoPlayer}, which is an open-source media player for Android, to interpret the EMPD manifest information taking into account the battery status and network bandwidth.

\noindent \textbf{Frame rate Scheduler (F-Scheduler):}
The source code for OpenH264 \cite{OpenH264} was modified to determine which frame rate is appropriate for each video chunk. The SAD values of the macroblocks are extracted when the H.264/AVC encoder performs the video compression process. These extracted SAD values of the macroblocks are utilized to analyze the motion variation of the video and to set the appropriate frame rates accordingly. 

\noindent \textbf{Video Processor (V-Processor):}
After the streaming server processes the uploaded video to generate videos with various resolutions and bitrates for the DASH service, V-Processor additionally processes the videos according to the schedule provided by F-Scheduler. This is achieved using the \textit{seeking} and \textit{concatenate} methods of FFmpeg \cite{FFmpeg} to generate videos with adaptive frame rates. The seeking method is used to split the video into multiple video chunks, and the frame rate of each video chunk is adjusted to match the schedule. Finally, the video chunks that have different frame rates are combined into a single video through the concatenate method.

Note that V-Processor is integrated into the H.264/AVC encoder for performance reasons. The implementation location of V-Processor can be changed depending on the deployment situation.

\section{Evaluation}\label{sec:evaluation}


\begin{table}
    \centering
   \resizebox{0.9\columnwidth}{!}{$
    \begin{tabular}{|| c | c | c | c | c || c ||} 
        \hline
        \thead{Video Name} & \thead{Length\\(min)} & \thead{Resolution\\(pixel)} & \thead{Bitrate\\(Kbps)} & \thead{Frame rate\\(fps)} & \thead{Type}\\ [0.5ex]
        \hline\hline
        \thead{\textbf{NBC} News Conference} & \thead{2.29} & \thead{1920x1080} & \thead{2130} & \thead{29.98} & 
        \multirow{5}{*}{\thead{A}} \\ 
        \hhline{-----|~||}
        \thead{\textbf{MIT} Course} & \thead{3.32} & \thead{1920x1080} & \thead{1191} & \thead{29.97} & \\
        \hhline{-----|~||}
        \thead{\textbf{Golf} | 2017 Back of\\Hope Founders Cup} & \thead{1.06} & \thead{1920x1080} & \thead{2256} & \thead{29.97} & \\
        \hline
        \thead{\textbf{PyeongChang} Olympic} & \thead{1.24} & \thead{1920x1080} & \thead{2908} & \thead{30.00} &
        \multirow{5}{*}{\thead{B}} \\
        \hhline{-----|~||}
        \thead{\textbf{Conan} Show} & \thead{3.1} & \thead{1920x1080} & \thead{3755} & \thead{29.98} & \\
        \hhline{-----|~||}
        \thead{\textbf{Kershaw} Baseball} & \thead{1.15} & \thead{1280x720} & \thead{1946} & \thead{29.97} & \\
        \hline
        \thead{\textbf{Pororo} Animation} & \thead{15.31} & \thead{1920x1080} & \thead{1085} & \thead{29.97} &
        \multirow{5}{*}{\thead{C}} \\
        \hhline{-----|~||}
        \thead{\textbf{Tennis} | Australian\\Open 2018} & \thead{10.04} & \thead{1280x720} & \thead{3075} & \thead{59.94} & \\
        \hhline{-----|~||}
        \thead{National \textbf{Geographic}} & \thead{13.53} & \thead{1280x720} & \thead{3115} & \thead{59.94} & \\
        \hline
    \end{tabular}
    $}
\caption{The characteristics of the videos used for the experiments. The bold word in the video name is used to represent that video in the experiments.}
\label{tab:video_characteristic}
\end{table}

\begin{table}
\centering
\resizebox{0.65\columnwidth}{!}{$
\begin{tabular}{lccc}
\toprule
Video & \makecell{EVSO} & \makecell{EVSO+} & \makecell{EVSO++} \\ \toprule
\textbf{NBC} & 24.82 & 22.44 & 19.18 \\
\textbf{MIT} & 19.35 & 16.51 & 14.27 \\ 
\textbf{Golf} & 23.47 & 21.41 & 18.86 \\ \toprule
\textbf{PyeongChang} & 27.64 & 26.78 & 24.06 \\
\textbf{Conan} & 25.95 & 24.46 & 21.84 \\
\textbf{Kershaw} & 26.36 & 24.44 & 21.64 \\ \toprule
\textbf{Pororo} & 25.51 & 23.75 & 20.83 \\
\textbf{Tennis} & 50.94 & 47.82 & 42.17 \\
\textbf{Geographic} & 53.38 & 49.44 & 42.63 \\
 \bottomrule
\end{tabular}
$}
\caption{The average frame rate of the videos processed through EVSO.}
\label{tab:reduced_fps}
\vspace{-1em}
\end{table}
\begin{table*}
\centering
\resizebox{0.85\textwidth}{!}{$    
\begin{tabular}{@{}lcccccccccccc@{}}
\toprule
\multirow{2}{*}{Video} & \multicolumn{3}{c}{\begin{tabular}[c]{@{}c@{}}\makecell{EVSO}\end{tabular}} & \multicolumn{3}{c}{\begin{tabular}[c]{@{}c@{}}\makecell{EVSO+}\end{tabular}} & \multicolumn{3}{c}{\begin{tabular}[c]{@{}c@{}}\makecell{EVSO++}\end{tabular}} & \multicolumn{3}{c}{\begin{tabular}[c]{@{}c@{}} 2/3 FPS\end{tabular}} \\ \cmidrule(lr){2-4} \cmidrule(lr){5-7} \cmidrule(lr){8-10} \cmidrule(l){11-13} 
 & SSIM & VQM & VMAF & SSIM & VQM & VMAF & SSIM & VQM & VMAF & SSIM & VQM & VMAF \\ \toprule
\textbf{NBC} & 99.53\% & 0.730 & 99.61\% & 99.41\% & 0.814 & 99.33\% & 99.06\% & 0.998 & 98.92\% & 98.52\% & 1.137 & 98.42\% \\
\textbf{MIT} & 99.68\% & 0.235 & 99.30\% & 99.63\% & 0.256 & 99.05\% & 99.56\% & 0.289 & 98.69\% & 99.52\% & 0.318 & 98.65\% \\ 
\textbf{Golf} & 98.64\% & 0.515 & 98.66\% & 98.50\% & 0.563 & 98.26\% & 98.03\% & 0.711 & 97.09\% & 96.95\% & 0.925 & 94.91\% \\ \toprule
\textbf{PyeongChang} & 99.29\% & 0.366 & 99.27\% & 99.12\% & 0.431 & 98.77\% & 97.87\% & 0.847 & 95.63\% & 94.20\% & 1.527 & 85.57\% \\
\textbf{Conan} & 99.04\% & 0.818 & 98.77\% & 98.76\% & 0.942 & 97.76\% & 98.20\% & 1.106 & 96.46\% & 96.66\% & 1.513 & 92.82\% \\
\textbf{Kershaw} & 98.54\% & 0.707 & 98.81\% & 98.18\% & 0.839 & 98.20\% & 96.60\% & 1.324 & 95.72\% & 93.57\% & 1.866 & 91.01\% \\ \toprule
\textbf{Pororo} & 98.92\% & 0.630 & 97.27\% & 98.50\% & 0.780 & 96.07\% & 97.62\% & 1.068 & 93.33\% & 95.94\% & 1.439 & 90.47\% \\
\textbf{Tennis} & 98.90\% & 0.806 & 99.02\% & 98.76\% & 0.892 & 98.62\% & 98.42\% & 1.051 & 97.29\% & 97.93\% & 1.249 & 95.32\% \\
\textbf{Geographic} & 99.12\% & 0.860 & 99.34\% & 98.94\% & 0.962 & 98.80\% & 98.73\% & 1.095 & 97.99\% & 98.45\% & 1.199 & 97.16\% \\
 \bottomrule
\end{tabular}
$}
\caption{Video quality scores according to the metrics of SSIM, VQM, and VMAF. SSIM and VMAF are expressed in percentages.}
\label{tab:quality_assessment}
\vspace{-1.0em}
\end{table*}

\begin{figure*}[!t]
    \includegraphics[width=0.95\textwidth]{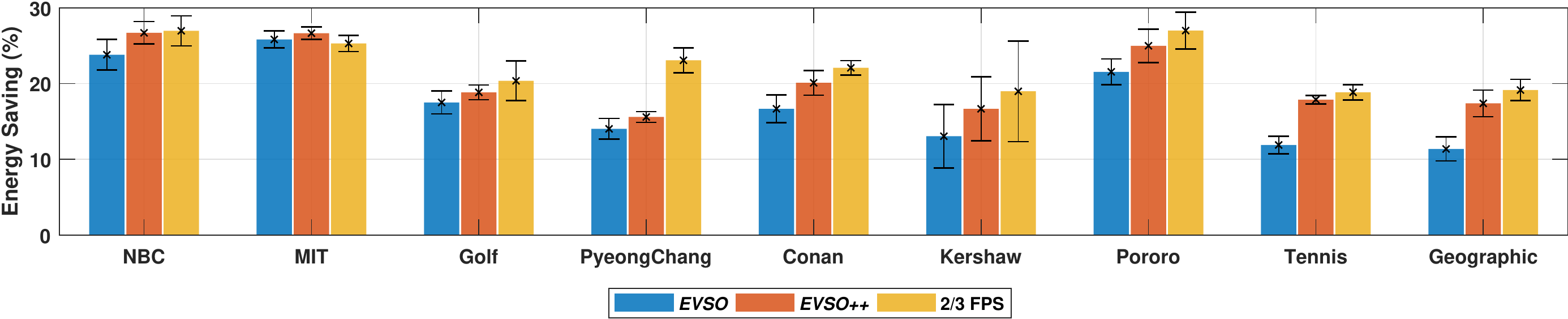}
    \caption{The percentage of energy savings while watching each video by streaming.}
    \label{fig:energy_saving}
    \vspace{-1.0em}
\end{figure*}
We evaluated the performance of the EVSO system by determining 1) how much the frame rate can be reduced; 2) how well the quality of the processed videos is maintained; 3) how much total energy can be saved; 4) how much do the processed videos affect the user experience; and 5) how much overhead is caused by EVSO.

\noindent \textbf{Videos used for experiments:} EVSO was evaluated based on nine videos of various categories such as sports, talk shows, lectures, etc. (available at \cite{ExperimentVideos}). As shown in Table~\ref{tab:video_characteristic}, these videos have various frame rates and resolutions in order to represent an environment similar to an actual streaming service. In addition, the videos are grouped into three types according to the degree of motion intensity: (A) Static, (B) Dynamic, and (C) Hybrid groups. The static group consists of videos that display almost identical scenes, such as lectures. The dynamic group consists of videos with a high level of motion intensity, such as sports. The hybrid group consists of videos that have both attributes of the two aforementioned groups. Unless otherwise noted, we conducted experiments on the videos processed to play for up to three minutes.

\noindent \textbf{Four different settings used for experiments (EVSO, EVSO+, EVSO++, 2/3 FPS):}
We configured EPF with three different settings according to the user's battery status, i.e., EVSO, EVSO+, and EVSO++. The worse the battery condition, the more aggressive EVSO is used (denoted by more + signs), allowing for a longer battery lifetime. For this reason, through multiple experiments, the values of \(s_{1}\), \(s_{2}\), \(s_{3}\), \(s_{4}\), and \(s_{5}\) in Equation~(\ref{eq:EPF}) are set to 0.6, 0.83, 0.9, 0.93, and 1 for EVSO, 0.5, 0.73, 0.83, 0.9, and 1 for EVSO+, 0.43, 0.6, 0.7, 0.8, and 0.93 for EVSO++, respectively. As a comparison, we also created an experimental group called 2/3 FPS that naively reduces the frame rate of the original video to two-thirds.

\noindent \textbf{Videos processed through EVSO:}
Table~\ref{tab:reduced_fps} shows the average frame rates of the videos processed according to battery levels. Since EVSO adaptively reduces the frame rate in consideration of the characteristics of the video, the resulting frame rate is quite different for each video. For example, the \textbf{MIT} video has little change in motion and is therefore reduced by more than one-third of the original frame rate on EVSO. On the other hand, the \textbf{Geographic} video has a relatively fast-changing motion, so EVSO reduces the frame rate of the video to only about five-sixths of the original frame rate.

\subsection{Video Quality Assessment}
We used three objective quality metrics to measure how much EVSO affected the quality of the videos: SSIM \cite{SSIM}, Video Quality Metric (VQM) \cite{VQM}, and Video Multimethod Assessment Fusion (VMAF) \cite{VMAF}. SSIM is an appropriate metric for calculating similarity based on images; however, because our experiments focused on videos, we also used VQM and VMAF. These metrics are more suited to measuring the subjective quality of the video than SSIM. A low value of VQM indicates a high quality video. On the other hand, the VMAF index has a range from 0 to 1, and a high value of VMAF indicates a high quality video. 

Table~\ref{tab:quality_assessment} shows the quality of the videos processed through EVSO compared to 2/3 FPS. As can be seen, EVSO provides better overall video quality than 2/3 FPS in all video cases. For the static group, both 2/3 FPS and EVSO maintain high quality because the motion intensity in the videos is comparatively low. The most prominent gap between EVSO and 2/3 FPS occurs in the dynamic group, where 2/3 FPS causes the VMAF metric to be less than 90\% indicating a severe degradation of the user experience. In addition, in the \textbf{Kershaw} video, the average frame rates of EVSO++ and 2/3 FPS are similar at 21.66 and 20, respectively, whereas the VMAF metric of EVSO++ is measured to be 5\% higher than 2/3 FPS. This is because EVSO adjusts the frame rate by analyzing each part of the video so that the video quality can be kept much higher than 2/3 FPS, which naively reduces the frame rate. 

\subsection{Energy Saving}
We used a Monsoon Power Monitor \cite{Monsoon} to measure the energy consumption of processed videos on a LG Nexus 5. The brightness of the smartphone was configured to 30\%, and the airplane mode was activated to reduce the effect from external variables, and Wi-Fi turned on to stream videos from the streaming server. To prevent thermal throttling caused by overheating, the temperature of the smartphone was cooled before each experiment.

We evaluated the energy consumption of nine videos in four different settings: Baseline, EVSO, EVSO++, and 2/3 FPS. Baseline is an original video and serves as a control group. EVSO, EVSO++, and 2/3 FPS use the same settings as described above. For accurate experiments, the energy consumption of each video was measured five times and then averaged.

Figure~\ref{fig:energy_saving} shows the ratio of energy savings when comparing Baseline to EVSO, EVSO++, and 2/3 FPS. Because EVSO adjusts the frame rate according to the motion intensity, the amount of energy saved varies considerably depending on the video characteristics. For example, the energy used by the \textbf{MIT} video in EVSO was reduced by about 27\%, but for the \textbf{Tennis} video, the reduction was only about 11\%. However, if the current battery status is bad, EVSO++ can be used to reduce the energy requirement similarly to the 2/3 FPS group, except for the \textbf{PyeongChang} video, with a reduction on average of 22\% compared to Baseline.

\begin{figure*}[!t]
\begin{subfigure} {0.33\textwidth}
    \includegraphics[width=1\textwidth]{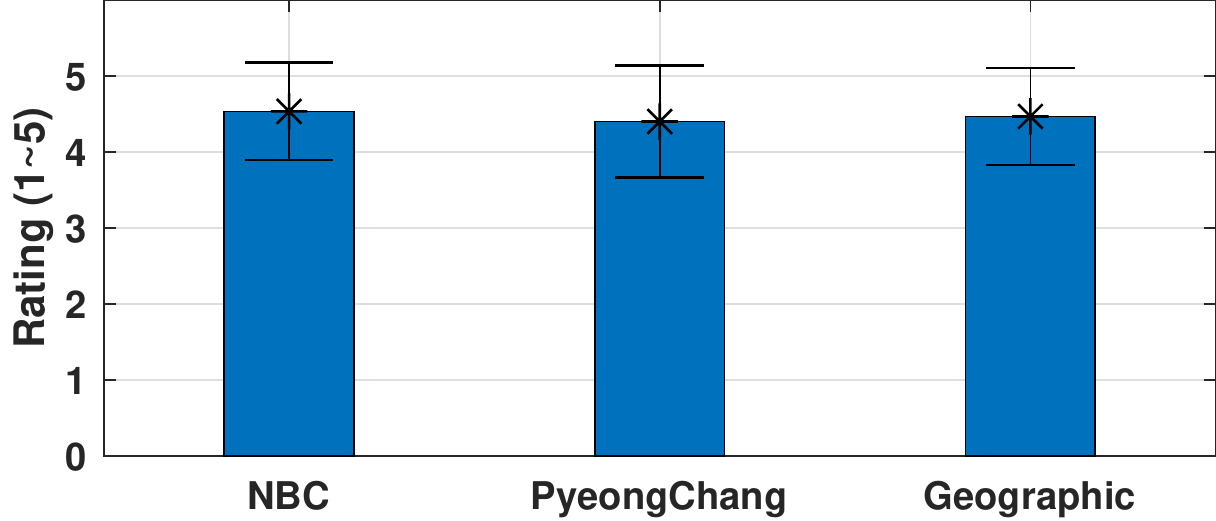}
    \caption{Quality rating from DSIS}
    \label{fig:DSIS}
\end{subfigure}
\begin{subfigure} {0.33\textwidth}
    \includegraphics[width=1\textwidth]{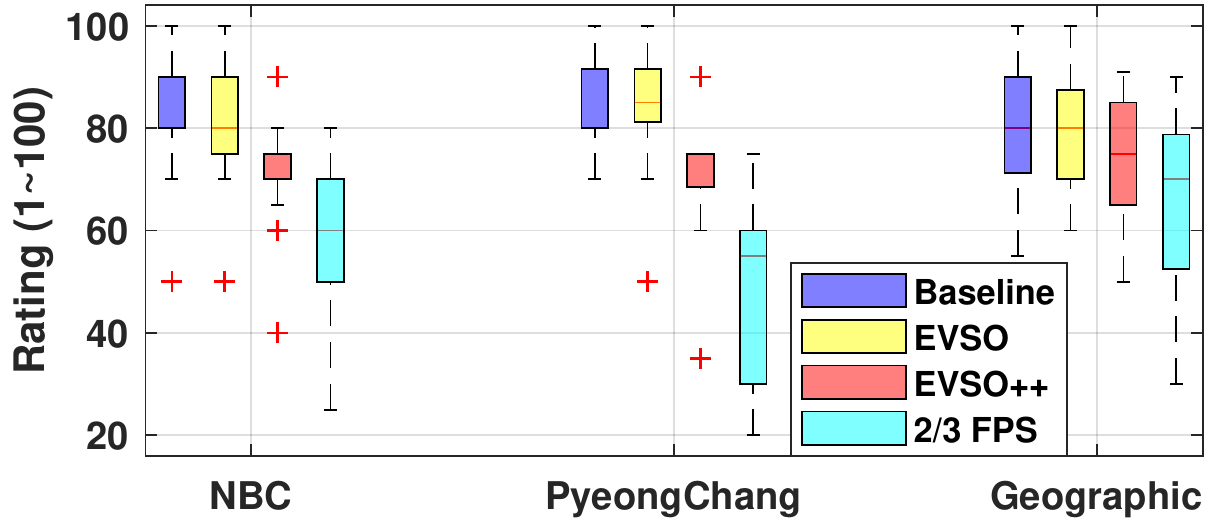}
    \caption{Quality rating from DSCQS}
    \label{fig:DSCQS_rating}
\end{subfigure}
\begin{subfigure} {0.33\textwidth}
    \includegraphics[width=1\textwidth]{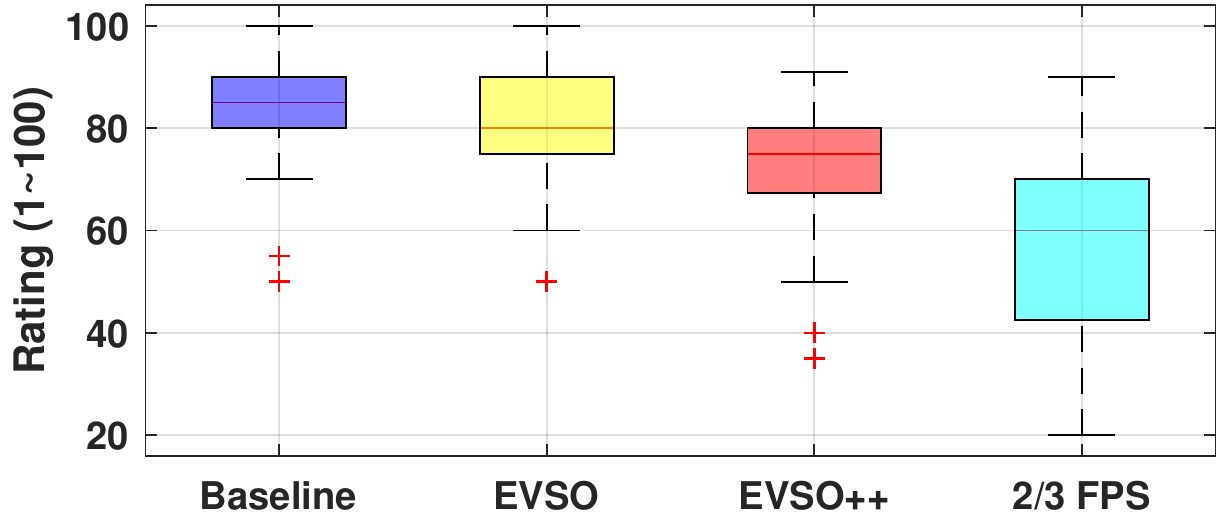}
    \caption{Quality rating distribution from DSCQS}
    \label{fig:DSCQS_distribution}
\end{subfigure}
\caption{Results of the user study through DSIS and DSCQS.}
\label{fig:user_study}
\vspace{-1.0em}
\end{figure*}

\subsection{User Study}
To analyze how the videos processed through EVSO affect the user experience, we recruited 15 participants (10 males) for the user study. We adopted the Double Stimulus Impairment Scale (DSIS) and the Double Stimulus Continuous Quality Scale (DSCQS), which are widely used for subjective quality assessments of systems \cite{kim2016content, RAVEN}. For the sake of experimentation, participants were allowed to watch one minute per video.

\noindent \textbf{DSIS:}
DSIS allows the participants to sequentially view two stimuli, i.e., the source video and the processed video, to assess the degree of impairment to the processed video. The evaluation index of DSIS is a five-point impairment scale of two stimuli: imperceptible (5 points), perceptible, but not annoying (4 points), slightly annoying (3 points), annoying (2 points), and very annoying (1 points). In DSIS, the participants are informed about which videos were original or processed.

We measured how much impairment occurred on the videos processed through EVSO. The participants were asked to watch three video types: \textbf{NBC} (Static), \textbf{PyeongChang} (Dynamic), and \textbf{Geographic} (Hybrid). The participants were informed in advance of which videos were processed, and they saw the original video and then the processed video in sequence.

Figure~\ref{fig:user_study}(a) shows the impairment rating of the videos. Most participants scored the processed videos as either "imperceptible" or "perceptible, but not annoying". This indicates that most participants did not recognize the difference between the original video and the processed video. In addition, the fact that all three different types of the video received high scores indicates that EVSO properly considers the motion intensity characteristics of the videos.

\noindent \textbf{DSCQS:}
DSCQS allows the participants to view two stimuli, i.e., the source video and the processed video, in random order to evaluate the differences in perceived visual quality. The evaluation index of DSCQS is a [0-100] scale for each stimulus. Unlike DSIS, DSCQS does not inform the participants about the setting of each video.

The participants were asked to watch the same three videos as in the DSIS method. For each video, the participants watched the videos using four different settings in random order and evaluated the quality of each video. The participants were allowed to watch the previously viewed videos again if they wanted.
Figures~\ref{fig:user_study}(b) and (c) show the quality rating and quality rating distribution using DSCQS. The participants could not distinguish between the original video (i.e., Baseline) and the processed video (i.e., EVSO). On the other hand, the participants were able to clearly discriminate 2/3 FPS, whose frame rate was naively reduced to two thirds. In addition, EVSO++, which aggressively reduces the frame rate, has an average frame rate similar to 2/3 FPS, but the average quality score is 16 points higher. This indicates that 
reducing the frame rate using EVSO leads to better quality than naively reducing the frame rate.

\subsection{System Overhead}
We measured the system overhead on a server equipped with 2.20 GHz\(\times\)40 processors and 135 GB of memory. Experiments were performed with the \textbf{NBC} video and the average was measured after repeating the session five times for accuracy.

\noindent \textbf{Scheduling overhead:}
We measured how much scheduling overhead occurred when the \textit{Frame rate Scheduler} (F-Scheduler) was built into the OpenH264 encoder. The OpenH264 encoder without and with F-Scheduler takes about 21.85 seconds and 22.02 seconds on average, respectively. The processing overhead incurred by F-Scheduler to schedule an adaptive frame rate for each video chunk is only 0.76\% on average, which clearly indicates that there is little or no overhead caused by the addition of F-Scheduler.
 
\noindent \textbf{Video processing overhead:}
We measured how much processing overhead occurred when the \textit{Video Processor} (V-Processor) processed a video based on the scheduling results from F-Scheduler. A video streaming server without V-Processor processes an uploaded video with various resolutions and bitrates, such as 720p (5 Mbps) and 480p (2.5 Mbps), to provide DASH functionality \cite{DASH}. In addition to this task, V-Processor processes videos according to three battery levels. The video streaming server without and with V-Processor takes about 143 seconds and 289 seconds on average, respectively. This indicates that the processing overhead caused by V-Processor is approximately 100\%. However, this video processing is internally handled by the streaming server and is a preliminary task performed prior to streaming the video to users. In addition, the time overhead can be significantly reduced if V-Processor is performed based on a selective strategy, such as processing only popular videos. Therefore, the overhead caused by V-Processor can be tolerated by both the user and streaming server.

\noindent \textbf{Storage overhead:}
Because the EVSO system generates additional videos based on the three battery levels, the amount of videos that need to be stored is nearly three times more than a conventional streaming server. This may be a strain on the storage requirement of the streaming server; however, this problem can also be alleviated if V-Processor is selectively performed only on popular videos, as described above. 

\section{Related Work}\label{sec:related}

\subsection{Energy Saving when Streaming Videos}
Lim et al. \cite{lim2016adaptive} extended the H.264/AVC encoder to provide a frame-skipping scheme during compression and transmission for mobile screen sharing applications. This scheme can reduce the energy consumption of mobile devices without significantly affecting the user experience, as similar frames are skipped during compression and transmission. In addition, this scheme does not transmit unnecessary frames to the mobile device because it works on the server side, thus there is a benefit of increased available network bandwidth on the client side. However, this scheme incurs high computational overhead because it calculates the similarity of all frames with the SSIM method \cite{RAVEN}. On the other hand, EVSO proposes a novel M-Diff method specialized for a video encoder and provides a flexible way to adjust the frame rate according to the current battery status of the mobile device.

Since the wireless interface consumes a considerable amount of energy on the mobile device, there have been several efforts to reduce the energy consumption by utilizing a playback buffer when streaming videos \cite{rao2011network, hu2015energy}. These approaches download a large amount of data in advance and store it in the player's playback buffer to increase the idle time of the wireless interface. These approaches are interoperable with EVSO. However, if users frequently skip or quit while watching the video, network bandwidth can be wasted because the previously downloaded data is no longer used.

Kim et al. \cite{kim2016content} adjusted a refresh rate of the screen by defining a content rate, which is the number of meaningful frames per second. If the content rate is low, this scheme reduces the refresh rate, which reduces the energy requirement without significantly affecting the user experience. However, the energy usage of the wireless interface, which is the main energy-consuming factor in a video streaming, cannot be reduced at all because only the refresh rate on the user side is adjusted.

\subsection{Energy Saving for Mobile Games}
Hwang et al. \cite{RAVEN} introduced a rate-scaling technique called RAVEN that skips similar frames in the rendering loop while playing games. This system uses Y-Diff method because the SSIM method has high computational overhead. The authors showed that Y-Diff has accuracy comparable to SSIM in human visual perception. On the other hand, EVSO proposes a novel M-Diff method specialized for a video streaming service. Moreover, RAVEN cannot be applied to a video streaming service because it performs rate-scaling in the rendering loop for mobile games.

Chen et al. \cite{chen2014fingershadow} proposed a method called FingerShadow that reduces the power consumption of the mobile device by applying a local dimming technique to the screen areas covered by user's finger while playing a game. However, this approach requires external interaction with user behavior. In contrast, EVSO is independent of external factors and performs power savings using the motion intensity information of the video.


\section{Conclusion and Future Work}\label{sec:conclusion}

In this paper, we propose EVSO, which is a system that opportunistically applies adaptive frame rates for videos without requiring any user effort. We introduce a similarity calculation method specialized for a video encoder and present a novel scheduling technique that assigns an appropriate frame rate to each video chunk. Our various experiments show that EVSO reduces the power requirement by as much as 27\% on mobile devices, with a little degradation of the user experience.

As future work, we plan to optimize various factors used in EVSO. A potential direction would be to use reinforcement learning to optimize decisions and continue learning through the policy training. The other direction would be to include more videos with various frame rates in experiments. 

\section*{Acknowledgment}
We thank the anonymous reviewers as well as our colleague Jaehyun Park for their valuable comments and suggestions. This research was a part of the project titled \enquote{SMART-Navigation project,} funded by the Ministry of Oceans and Fisheries, Korea. This research was also supported by the National Research Foundation of Korea (NRF) grant funded by the Korean government (MSIP) (No. 2017R1A2B4005865).



\bibliographystyle{IEEEtran}
\bibliography{bibliography}
%
%
%

\end{document}